\documentclass[final,1p,times,authoryear]{elsarticle}
\usepackage{amssymb}
\usepackage{lineno}

\usepackage{rotating}
\usepackage{color}

\journal{Icarus}

\begin{document}

\begin{frontmatter}



\title{New Insights on Jupiter's Deep Water Abundance from Disequilibrium Species}


\author{Dong Wang}
\ead{dw459@cornell.edu}
\address{Department of Astronomy, 610 Space Sciences Building, Cornell University,
    Ithaca, NY 14853}

\author{Peter J. Gierasch}
\address{Department of Astronomy, 610 Space Sciences Building, Cornell University,
    Ithaca, NY 14853}

\author{Jonathan I. Lunine}
\ead{jlunine@astro.cornell.edu}
\address{Center for Radiophysics and Space Research, Space Sciences Building, Cornell University,  Ithaca, NY 14853, USA}

\author{Olivier Mousis}
\address{Department of Astronomy, 610 Space Sciences Building, Cornell University, Ithaca, NY 14853}
\address{Aix Marseille Universit\'e, CNRS, LAM (Laboratoire d'Astrophysique de Marseille) UMR 7326, 13388, Marseille, France}

\begin{abstract}

The bulk water abundance on Jupiter potentially constrains the planet's formation conditions. We improve the chemical constraints on Jupiter's deep water abundance in this paper. The eddy diffusion coefficient is used to model vertical mixing in planetary atmosphere, and based on laboratory studies dedicated to turbulent rotating convection, we propose a new formulation of the eddy diffusion coefficient for the troposphere of giant planets. The new formulation predicts a smooth transition from the slow rotation regime (near the equator) to the rapid rotation regime (near the pole). We estimate an uncertainty for the newly derived coefficient of less than 25$\%$, which is much better than the one order of magnitude uncertainty used in the literature. We then reevaluate the water constraint provided by CO, using the newer eddy diffusion coefficient. We considered two updated CO kinetic models, one model constrains the water enrichment (relative to solar) between 0.1 and 0.75, while the other constrains the water enrichment between 3 and 11. \end{abstract}


\end{frontmatter}


\section{Introduction}
\label{intro}

The bulk abundances of oxygen in Jupiter and Saturn potentially constrain conditions in the Sun's protoplanetary disk.  However, determining these abundances through the direct measurement of water, the dominant carrier of oxygen in the envelopes of these objects,  is very difficult. Galileo probe measurements show the effect of dynamical processes on the water abundance down to 22 bars (Wong et al. 2004), while ground-based microwave observations are not sufficiently sensitive to provide a deep water abundance (that is, below the meteorological layer) for either body (de Pater \& Massie 1985). A determination of the deep ($>$ 50 bar) water abundance on Jupiter should be obtained by the microwave radiometer aboard the Juno spacecraft set to arrive at Jupiter in 2016 (Janssen et al. 2005; Helled \& Lunine 2014). There is no similar possibility for Saturn in the near future because, even though the Cassini spacecraft will be put in a Juno-like orbit in 2017, it does not carry a microwave radiometer.

An alternative way to determine water abundance, through disequilibrium species observed in Jupiter and Saturn's troposphere, is a long-standing approach that goes back to Prinn \& Barshay (1977) (see Visscher \& Moses 2011 for an extensive list of published papers on this subject). The abundance of disequilibrium species depends on the relevant chemical kinetics, which determines the chemical loss rate, and the eddy diffusion coefficient, which determines the efficiency of vertical mixing. 
Our study is timely, in spite of the long history of published papers, for three reasons. 
First, we derived a new formulation of the eddy diffusion coefficient based on laboratory studies of turbulent rotating convection. The new formulation systematically describes the transition from slow-rotation convection to rapid-rotation convection with significantly less uncertainty than previously. 
Secondly, we used the two most updated CO kinetic models to place constraints on Jupiter's deep water abundance.  
Third, a possible future mission to deploy a descent probe into Saturn's atmosphere, if conducted, will almost certainly be a ``New Frontiers" medium-class mission (National Research Council 2011), or an ESA M-class mission (Mousis et al. 2014). Such a probe will probably not be able to get to the base of the water cloud which is essential to determining directly the deep oxygen abundance on Saturn. Indirect methods including using disequilibrium species as described here may be the only way to determine oxygen abundance even through probe measurements, and therefore a study is warranted using the most recent kinetics to assess whether such an approach provides a well-constrained oxygen value. Our results identify and quantify significant ambiguities inherent in such an approach. 

The paper is organized as follows. In section 2, we analyze the results from rotating tank experiments and propose a new formulation of eddy diffusion coefficient. In section 3, we derive constraints on the deep water abundance from CO measurements with the kinetic information from two different models. In section 4, we discuss the implication on Jupiter's formation and potential improvements relative to the current model.  

\section{A New Formulation for the Deep Eddy Diffusion Coefficient}
\label{K_eddy}

In the atmosphere of Jupiter, heat is assumed to be transported by vertical eddy diffusion. The eddy diffusion coefficient $K_{\rm eddy}$ is introduced to measure the efficiency of vertical diffusion. In the convective part of the atmosphere, the heat flux and superadiabatic temperature gradient can be related to $K_{\rm eddy}$ by the following equation: 
\begin{equation}\label{K_eddy_def}
F = - \rho c_p K_{\rm eddy}  (\frac{dT}{dr}-{\frac{dT}{dr}}|_{ad}),
\end{equation} 
where $F$ is the internal heat flux, $\rho$ is the mass density, $c_p$ is the specific heat per unit mass, and $dT/dr -  dT/dr|_{ad}$ is the superadiabatic temperature gradient.  
Formulations of $K_{\rm eddy}$ in terms of heat flux $F$, rotation rate $\Omega$ and fluid thermal properties are derived based on mixing length theory or perturbation of linearized equations (Stone 1976; Flasar \& Gierasch 1978; Stevenson 1979), predicting $K_{\rm eddy}$ near CO quench level ($\sim$ 1000 K, 300 bars) for Jupiter to be between 1$\times$10$^7$ cm$^2$ s$^{-1}$ and  
1$\times$10$^9$ cm$^2$ s$^{-1}$ (e.g. B\'{e}zard et al 2002; Visscher et al. 2010), and this value is widely used in theoretical modeling of disequilibrium chemistry. One difficulty in improving the estimation is the lack of observation. No natural convective system under rapid rotation, like the interior of giant planets or the Earth core, can be easily observed. However, the estimation of $K_{\rm eddy}$ could be improved by utilizing results from laboratory studies on turbulent rotating convection. Laboratory studies on turbulent rotating convection have been done since 1980s, however, application to giant planet convection has hitherto been limited. Here, we  summarize relevant results of these laboratory studies, and propose a new formulation of $K_{\rm eddy}$.   

In section 2.1, we review theoretical investigations on $K_{\rm eddy}$. In section 2.2, we summarize results from rotating tank experiments, and present the new formulation for $K_{\rm eddy}$. In section 2.3, we apply the new formulations to Jupiter and Saturn, and predict $K_{\rm eddy}$ profiles for these two planets.  

\subsection{Theory on Eddy Diffusion Coefficient}

By analogy to molecular diffusion coefficient, $K_{\rm eddy}$ can be approximated as the product of vertical convective velocity $w$ and a mixing length $l$, representing a typical distance a parcel could travel before it lost its identity. 
Therefore, equation (\ref{K_eddy_def}) can be rearranged as 
\begin{equation}\label{MLT}
F \sim -\rho c_p w \delta T,
\end{equation}
where $\delta T$ = $(dT/dr - dT/dr|_{ad})l$ is the temperature fluctuation. 
A parcel's kinetic energy is obtained from the work done by buoyancy force over a mixing length $l$, thus
\begin{equation}\label{equ: v}
w^2 \sim -\alpha g \delta T l,
\end{equation}
where $\alpha$ is the thermal expansion coefficient and $g$ is the acceleration of gravity. 
With equation (\ref{MLT}) and (\ref{equ: v}), we find the convective velocity 
\begin{equation}\label{equ: w_stone}
w \sim (\frac{\alpha g F}{\rho c_p }l)^{1/3}. 
\end{equation}
The mixing length is usually assumed to be a pressure scale height $H$, thus eddy diffusion coefficient can be estimated as (Stone 1976)
\begin{equation}\label{K_stone}
K_{\rm eddy} \sim w l \sim (\frac{\alpha g F}{\rho c_p }H)^{1/3}H,
\end{equation}
Stone (1976)'s estimation ignored the effect of rotation on $K_{\rm eddy}$, however, rotation could have an important effect on convection in suppressing vertical mixing (Guillot et al. 2004). The importance of rotation can be measured by a Rossby number
\begin{equation}\label{Ro}
Ro = \frac{v}{fl},
\end{equation}
where $f = 2\Omega sin\phi$ is the Coriolis parameter, and $\phi$ is the latitude. The Rossby number is defined as the ratio of inertial to Coriolis force, therefore, lower $Ro$ means Coriolis acceleration is more important. Near the CO quench level, we find $Ro \approx 0.01/(sin \phi)$. Therefore, near the equator, rotation has little effect, while at extra-equatorial latitudes, rotation is important in suppressing turbulent convection. 

The trend is consistent with Flasar \& Gierasch (1978)'s results. In the limit of rapid rotation, Flasar \& Gierasch (1978) analyzed the linear modes generated by the perturbation of a superadiabatic and inviscid fluid in plane geometry, and identified the most unstable modes that transport the most heat. Assuming shear instability limits the growth rate, they found 
\begin{equation}\label{K_fg}
K_{\rm eddy} \sim (\frac{\alpha g F}{\rho c_p})^{3/5} (\frac{H}{2\Omega sin\phi})^{4/5},
\end{equation}
and 
\begin{equation}\label{w_fg}
w \sim (\frac{\alpha g F}{\rho c_p})^{2/5} (\frac{H}{2\Omega sin\phi})^{1/5}
\end{equation}
at extra-equatorial regions, while near the equator, the formulation is the same as equations (\ref{equ: w_stone}) $\&$ (\ref{K_stone}).  
(Equations (\ref{K_fg}) and (\ref{w_fg}) are rearranged from equation (5.3) in Flasar \& Gierasch 1978)
  
Equations (\ref{K_stone}) and (\ref{K_fg}) are widely used in estimating $K_{\rm eddy}$ (e.g. B\'{e}zard et al. 2002; Visscher et al. 2010). 
In comparison to equations (\ref{K_stone}) and (\ref{K_fg}), laboratory experiments on turbulent rotating convection indicate the same scaling as equation (\ref{K_stone}) for slow rotation, but a different scaling from equation (\ref{K_fg}) for rapid rotation. We will discuss the new scalings from rotating tank experiments in section 2.2, but here we will show that the new scalings can be easily derived based on the assumption that overturning timescale is limited by the rotational timescale $\Omega^{-1}$, instead of $H/w$. With this assumption, the relevant length scale would be $l = w/\Omega$, and the velocity scale would be
\begin{equation}
w \sim (\frac{\alpha g F}{\rho c_p \Omega})^{1/2}, 
\end{equation}
according to equations (\ref{MLT}) and (\ref{equ: v}). This velocity scale was found to be consistent with the convective velocity data from a three dimensional anelastic simulation of the convective envelope of Jupiter (Showman et al. 2011). The relevant length scale would be 
\begin{equation}
l \sim w/\Omega \sim (\frac{\alpha g F}{\rho c_p \Omega^3})^{1/2}
\end{equation}
instead of a pressure scale height. Therefore, we can formulate a new scaling for the eddy diffusion coefficient using the velocity and length scale described here. The eddy diffusion coefficient would be 
\begin{equation}\label{K_eddy_scaling}
K_{\rm eddy} \sim wl \sim \frac{\alpha g F}{\rho c_p \Omega^2}.
\end{equation}
Here we neglected all the prefactors in the scalings, however, these prefactors can be determined from laboratory measurements.

\subsection{Scalings from Rotating Tank Experiments}\label{lab_result}

A set of scalings for integral length scale and convective (r.m.s.) velocity was proposed and validated by rotating tank experiments. The laboratory experiments (e.g. Boubnov \& Golitsyn 1986, 1990; Fernando et al. 1991; Maxworthy \& Narimousa 1994; Coates \& Ivey 1997) are done in a rotating tank filled with water while the tank can have rotation with respect to its vertical axis. The bottom of the tank is heated in order to generate thermal convection, while the upper surface is open. Flow speed and temperature are directly measured. It is observed that two physically distinct regimes are identified, depending on rotation rate: (1) fully developed turbulence when the rotation rate is small, and (2) geostrophic turbulence when the rotation rate is large. Two sets of scalings for convective velocity $w$ and integral length scale $l$ are also proposed for these two regimes, respectively.     
For fully developed turbulence, the convective velocity and integral length scale are (e.g. Adrian et al. 1986; Deardorff 1972; Fernando et al. 1991)
\begin{equation}\label{non-rotating}
w_{\rm norot} = \alpha (\frac{\alpha g F}{\rho c_p} l)^{1/3}, l_{\rm norot} = h, 
\end{equation} 
where $h$ is the height of the fluid in the tank. This expression offers experimental support for the estimation by Stone (1976), but only applies to fully developed turbulence with weak rotational effects. The coefficient $\alpha \approx 0.6$ is determined from experimental measurements (Adrian et al. 1986; Fernando et al. 1991). A different scaling, for geostrophic turbulence (Fernando et al. 1991; Maxworthy \& Narimousa 1994; Coates \& Ivey 1997; Fernando \& Smith 2001; Levy \& Fernando 2002), is given by 
\begin{equation}\label{rotating}
w_{\rm rot} = \beta_1 (\frac{\alpha g F}{\rho c_p \Omega})^{1/2}, l_{\rm rot} = \beta_2 (\frac{\alpha g F}{\rho c_p \Omega^3})^{1/2},
\end{equation} 
where $\beta_1 = 1.2\pm0.6$ (Fernando et al. 1991; Maxworthy \& Narimousa 1994; Coates \& Ivey 1997), and $\beta_2 \approx 1.1$ (Fernando et al. 1991). 

The transition between fully developed turbulence and geostrophic turbulence is governed by a natural Rossby number (e.g. Maxworthy \& Narimousa 1994; Jones \& Marshall 1993), defined as 
\begin{equation}\label{Ro_star}
Ro^{\star} = l_{\rm rot} / h =  (\frac{\alpha g F}{\rho c_p \Omega^3})^{1/2}\frac{1}{h}. 
\end{equation}
There exists a transitional natural Rossby number $Ro^{\star}_t$. When $Ro^{\star}$ $<$ $Ro^{\star}_t$, turbulent convection is strongly inhibited by rotation, integral length scale is smaller than mixed layer height, and rotating scaling (equation \ref{rotating}) applies; when $Ro^{\star}$ $>$ $Ro^{\star}_t$, turbulent convection is weakly affected by rotation, thus non-rotating scaling applies (equation \ref{non-rotating}). Based on velocity data, Coates \& Ivey (1997) found $0.057 \leq Ro^{\star}_t \leq 0.14$, and Cui \& Street (2001) found $0.042\leq Ro^{\star}_t \leq 0.12$. 

\begin{figure}
\begin{center}
\resizebox{\hsize}{!}{\includegraphics[angle=0]{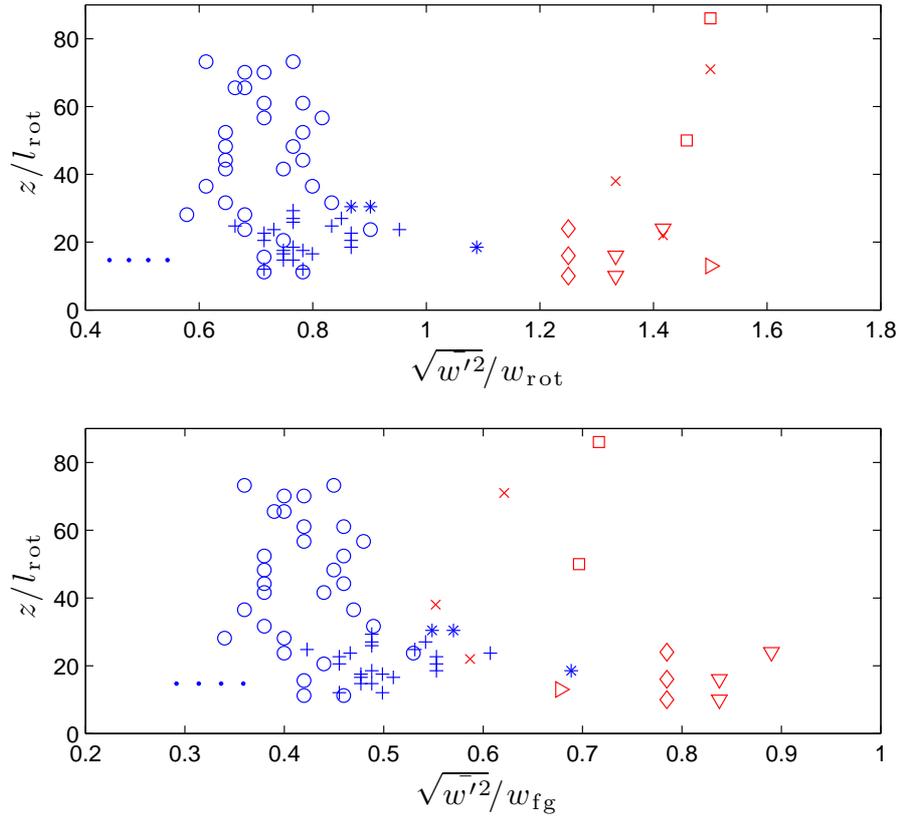}}
\caption{Normalization of vertical r.m.s. velocity $\sqrt{\overline{w'^2}}$ by $w_{\rm rot}$ and $w_{\rm fg}$, respectively. Vertical axis represents the height $z$ (relative to the bottom of the tank) where $\sqrt{\overline{w'^2}}$ is measured. The data colored in red are extracted from Fig 16 in Fernando et al. (1991) and the data colored in blue are extracted from Fig 6 in Coates \& Ivey (1997). Different marker types correspond to different sets of experiments, with different rotation rates and prescribed heat fluxes. We only plot data with $z/l_{\rm rot}$ $>$ 30 ($Ro^{\star}$ $<$ 0.03), because that is where we expect rotating scaling applies. The data points should fall along a vertical line if the scaling could represent $\sqrt{\overline{w'^2}}$ well.}
\label{collapse_data}
\end{center}
\end{figure}

One directly measurable quantity from these rotating tank experiments is the flow velocity. In Fig. \ref{collapse_data}, we normalize the measured r.m.s. vertical velocity $\sqrt{\overline{w'^2}}$ using $w_{\rm rot}$ and another velocity scaling 
\begin{equation}
w_{\rm fg} \sim (\frac{\alpha g F}{\rho c_p})^{2/5} (\frac{h}{2\Omega})^{1/5},
\end{equation} 
which is that derived by Flasar $\&$ Gierasch (1978). If the scalings $w_{\rm rot}$ or $w_{\rm fg}$ can represent the vertical r.m.s. velocity $\sqrt{\overline{w'^2}}$, we would expect $\sqrt{\overline{w'^2}}/w_{\rm rot}$ or $\sqrt{\overline{w'^2}}/w_{\rm fg}$ to be a constant. We use two experimental datasets for $\sqrt{\overline{w'^2}}$. One is from Fernando et al. (1991), and the other is from Coates \& Ivey (1997). In the figure, the dataset from Fernando et al. (1991) is indicated by blue markers, and the dataset from Coates $\&$ Ivey (1997) us indicated by red markers. In the upper plot of Fig. \ref{collapse_data}, data points are normalized by $w_{\rm rot}$. Both datasets are aligned at values that are similar but not identical. This indicates slightly different pre-factors before the scaling. We believe the difference is due to systematic error in measuring flow speed. In the lower plot, blue markers are well clustered, but red markers are subject to large scatter. Therefore, data from Fernando et al. (1991) are not well fitted by $w_{\rm fg}$. Overall, $w_{\rm rot}$ looks better in fitting experimental datasets than $w_{\rm fg}$, but the difference is not very significant.  

\begin{figure}
\begin{center}
\resizebox{\hsize}{!}{\includegraphics{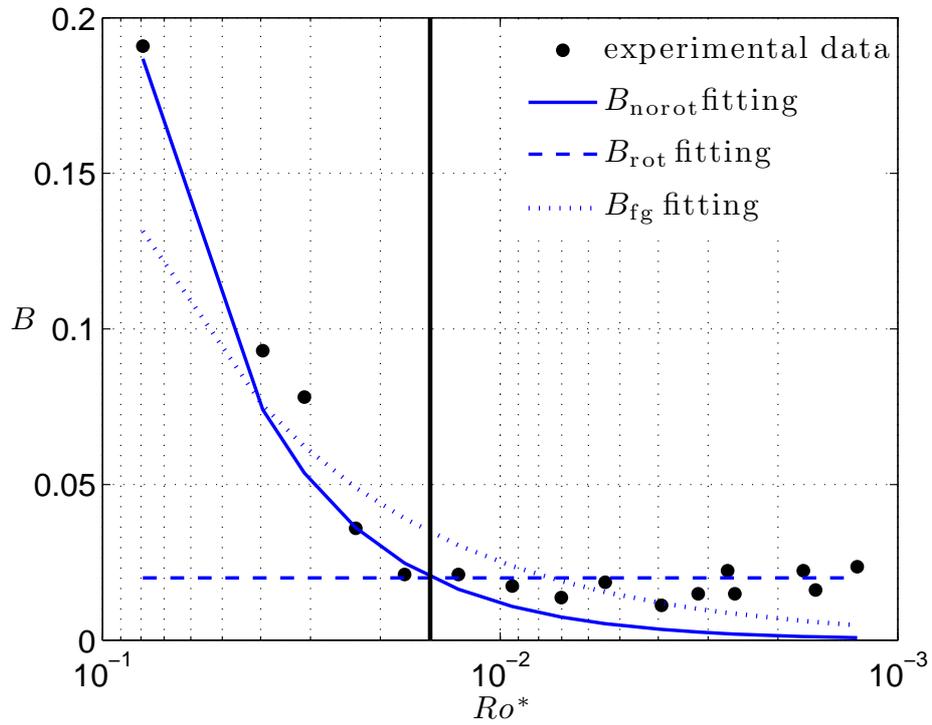}}
\caption{Scatterplot of B as a function of $Ro^{\star}$. $B = (\alpha g d\bar{T}/dz)/(\Omega^2)$, and $Ro^{\star} = l_{\rm rot}/h$. This figure is based on Fig 8 in Ivey $\&$ Fernando (2002). Black filled circles represent experimental data extracted from Fig 8 in Ivey $\&$ Fernando (2002). Solid curve is the fit using $B_{\rm norot}$ defined by equation (\ref{B_norot}), dashed line is the fit using $B_{\rm rot}$ defined by equation (\ref{B_rot}), and dotted curve is the fit using $B_{\rm fg}$ given by equation (\ref{B_fg}). The vertical line is  showing the transition from high $Ro^{\star}$ regime to low $Ro^{\star}$ regime, corresponding to $Ro^{\star}_{t}$ = 0.015. } 
\label{B_Ro_plot}
\end{center}
\end{figure}

Another measurable quantity is the vertical temperature gradient (e.g. $d\bar{T}/dz$). Vertical temperature gradient in non-rotating turbulent convection is almost negligible because vertical mixing is very strong and the fluid is very well mixed. However, in rotating fluid, because of the inhibition of vertical mixing by rotation, there are much larger temperature gradients. Now we define a dimensionless quantity $B$ to normalize $d\bar{T}/dz$, and its definition is 
\begin{equation}\label{B_def}
B = -\frac{\alpha g d\bar{T}/dz}{\Omega^2}.
\end{equation}
We derive the scalings for $B$ based on scalings for $w$ and $l$. For rapid rotation, $K_{\rm eddy, rot} = C_{\rm rot}^{-1} w_{\rm rot} l_{\rm rot}$, where $C_{\rm rot}$ is a constant, and $C_{\rm rot}^{-1}$ represents the correlation between $w_{\rm rot}$ and $l_{\rm rot}$. Using equation (\ref{rotating}) and (\ref{B_def}), we find $B_{\rm rot}$ can be expressed as
\begin{equation}\label{B_rot}
B_{\rm rot} = \frac{\alpha g F}{\rho c_p K_{\rm eddy, rot}\Omega^2} = C_{\rm rot}.
\end{equation} 
Another set of scalings (e.g., equation \ref{K_fg} $\&$ \ref{w_fg}) for rapid rotating convection is from Flasar \& Gierasch (1978). The corresponding scaling for the eddy diffusion coefficient is 
\begin{equation}\label{K_eddy_fg_exp}
K_{\rm eddy, fg} = C_{\rm fg}^{-1}(\frac{\alpha g F}{\rho c_p})^{3/5} (\frac{h}{\Omega})^{4/5},
\end{equation}
where $C_{\rm fg}$ is a constant.
Therefore, $B_{\rm fg}$ can be expressed as
\begin{equation}\label{B_fg}
B_{\rm fg} = \frac{\alpha g F}{\rho c_p K_{\rm eddy, fg}\Omega^2} = C_{\rm fg} Ro^{\star4/5}. 
\end{equation} 
Recall that $Ro^{\star}$ is the natural Rossby number defined by equation (\ref{Ro_star}). For slow rotation, using scalings from equation (\ref{non-rotating}), we find 
\begin{equation}\label{B_norot}
B_{\rm norot} = C_{\rm norot} Ro^{\star4/3}.
\end{equation} 
Both $B$ $\&$ $Ro^{\star}$ are measurable quantities in the experiments, therefore, datasets of ($Ro^{\star}$, $B$) are able to provide a test to the scalings derived above. In Fig \ref{B_Ro_plot}, we made a scatterplot of ($Ro^{\star}$, $B$) measured from experiments (Ivey \& Fernando et al. 2002). The scalings $B_{\rm rot}$, $B_{\rm norot}$ and $B_{\rm fg}$ as a function of $Ro^{\star}$ are overplotted on the same figure. Inspection of the fitting in Fig. \ref{B_Ro_plot} reveals that $B_{\rm rot}$ is clearly better in fitting data than $B_{\rm fg}$ for small $Ro^{\star}$ values, and $B_{\rm norot}$ can fit data well at large $Ro^{\star}$ values. In summary, the experimental data supported the scalings (\ref{non-rotating}), (\ref{rotating}), (\ref{B_rot}), $\&$ (\ref{B_norot}). The pre-factors are determined to be $C_{\rm norot} = 5.5\pm0.5$ and $C_{\rm rot} = 0.020 \pm 0.005$ (Ivey \& Fernando 2002).      

Now that we have experimentally validated formulations for $B$ or equivalently, $d\bar{T}/dz$, we determine the pre-factors in the scalings for $K_{\rm eddy}$: 
\begin{equation}\label{K_19}
K_{\rm eddy} = \frac{\alpha g F}{\rho c_p B_{\rm norot} \Omega^2} = (0.18\pm0.02)(\frac{\alpha g F}{\rho c_p})^{1/3} h^{4/3},\ \textrm{for} \ Ro^{\star} > Ro^{\star}_t;
\end{equation}
\begin{equation}\label{K_20}
K_{\rm eddy} = \frac{\alpha g F}{\rho c_p B_{\rm rot} \Omega^2} = (50\pm10)\frac{\alpha g F}{\rho c_p \Omega^2},\ \textrm{for} \ Ro^{\star} < Ro^{\star}_t.
\end{equation}
 $Ro^{\star}_t$ is identified in Fig \ref{B_Ro_plot} by locating the transition from $B_{\rm norot}$ scaling to $B_{\rm rot}$ scaling. $Ro^{\star}_t$ is found to be 0.015 with an uncertainty of 20$\%$ (Ievy $\&$ Fernando 2002). This value is consistent with estimations by Coates \& Ivey (1997) and Cui \& Street (2001) using velocity data. In the following sections, we will use $Ro^{\star}_t$ = 0.015.  

\subsection{Eddy Diffusion Coefficient in the Atmosphere of Jupiter and Saturn}
 
Now we have the scalings for $K_{\rm eddy}$ given by equation (\ref{K_19}) $\&$ (\ref{K_20}). An extrapolation is needed to apply this scaling to the atmospheres of giant planets because the flux Rayleigh number $Ra_f$ in the experiments cannot reach as high as that in giant planets. The flux Rayleigh number is defined as $Ra_f = \frac{\alpha g F H^4}{\rho c_p \kappa^2 \nu}$, where $\kappa$ is the thermal diffusivity and $\nu$ is molecular viscosity. It is defined as equivalent to the Rayleigh number $Ra = \frac{\alpha g dT/dz H^4}{\kappa \nu}$, but more straightforward than $Ra$ because the quantity $F$ is usually available rather than $dT/dz$ in astrophysical bodies. For Jupiter, $Ra_f$ is estimated to be about $10^{30}$ near $T = 1000 K$ using the thermal properties calculated in French et al. (2010). In the experiments, $Ra_f = 10^{12}\sim10^{13}$. Therefore, applying the experimentally derived scaling to Jupiter's atmosphere is an extrapolation, but it is likely that the extrapolation is reasonable for the following reasons. (1) It is generally believed that at very high $Ra_f$, molecular viscosity and diffusivity will not affect the property of turbulent transport. This happens when $Ra_f$ is much larger than the critical flux Rayleigh number $Ra_{f,c}$. From Fig. 4 in Fernando $\&$ Smith (2002), the $Ra_f$ implied in the experimental setup is at least three order of magnitude higher than the $Ra_{f,c}$, and most of them are four or five order of magnitude higher. Therefore, it is reasonable to assume the scalings are independent of molecular viscosities and diffusivities. (2) Non-rotating scaling (equation \ref{K_19}) is the same as the prediction by mixing length theory, which is another piece of evidence that the experiments have probed the highest $Ra_f$ regime. 

The experiments are set up in a plane parallel geometry and the rotation axis is aligned with the gravity, while Jupiter has a spherical geometry and locally the gravity and rotation vector is misaligned except at the pole. However, under some approximations, the experimental results are applicable to Jupiter's atmosphere as well. First, the vertical length scale ($\sim H$) is much smaller than the horizontal length scale ($\sim R_{\rm Jup}$), thus the curvature of the geometry is not important here. Second, since we assume heat is primarily transported by eddies, whose scale in the atmosphere is much smaller than planetary radii, $f$-approximation can be made in the governing Navier Stokes equations, and the dynamics thus only depend on the Coriolis parameter $f = 2\Omega sin \phi$, where $\phi$ is the latitude. Therefore, the scalings applicable to all latitudes can be derived by replacing $\Omega$ with $\Omega sin \phi$ in the equations (\ref{Ro_star}), (\ref{K_19}) $\&$ (\ref{K_20}). 
A transitional latitude $\phi_t$ exists, and can be related to $Ro^{\star}_t$ by  
\begin{equation}\label{phi_t}
sin\phi_t = (\frac{\alpha g F}{\rho c_p H^2 Ro^{\star 2}_t})^{1/3}\frac{1}{\Omega},
\end{equation}
$K_{\rm eddy}$ in the atmosphere thus can be written based on scalings given by equation (\ref{K_19}) $\&$ (\ref{K_20}):  
\begin{equation}\label{K_eddy_non}
K_{\rm eddy} = (0.18\pm0.02)(\frac{\alpha g F}{\rho c_p})^{1/3} H^{4/3},\textrm{ for } \phi < \phi_t,
\end{equation} 
\begin{equation}\label{K_eddy_rot}
K_{\rm eddy} = (50\pm10)\frac{\alpha g F}{\rho c_p (\Omega sin\phi)^2}, \textrm{ for } \phi > \phi_t,
\end{equation} 
while $\phi_t$ can be determined locally based on equation (\ref{phi_t}) with $Ro^{\star}_t$ = 0.015. Clearly $\phi_t$ is a function of altitude. At the level where T $\sim$ 1000 K (near the quench level), we find $\phi_t = 19^{\circ}$ for Jupiter, and $\phi_t = 5^{\circ}$ for Saturn.

The adiabatic profile for Jupiter and Saturn are calculated stepwise following the method described in Fegley $\&$ Prinn (1985). For Jupiter, we use $T$ = 427.71 K at $P$ = 22 bars as our reference point (Seiff et al. 1998). The compositions considered is $X_{\rm H_2}$ = 0.864 and $X_{\rm He}$ = 0.136 (Niemann et al. 1998). For Saturn, we use $T$ = 134.8 K at $P$ = 1 bar (Lindal et al. 1985) as our reference point, and the composition considered are $X_{\rm H_2}$ = 0.881 and $X_{\rm He}$ = 0.119 (Conrath $\&$ Gautier 2000). Strictly speaking, a wet adiabat would be more appropriate within the water cloud. However, since the quench level is around 400 bars, the wet adiabat is only a small portion of the extrapolated regions. Although we expect the use of a wet adiabat would yield a more accurate adiabatic profile, the difference from our calculated dry adiabat would be small. Along the calculated adiabatic profile, we calculate quantities such as $\alpha$, $\rho$, and pressure scale height $H$ using the ideal gas law since the part of atmosphere we considered is close to ideal gas. The internal heat flux for Jupiter is estimated to be 5.444 $\pm$ 0.425 W m$^{-2}$ (Hanel et al. 1981), and for Saturn it is estimated to be 2.01 $\pm$ 0.14 W m$^{-2}$ (Hanel et al. 1983).   

\begin{figure}
\begin{center}
\resizebox{\hsize}{!}{\includegraphics{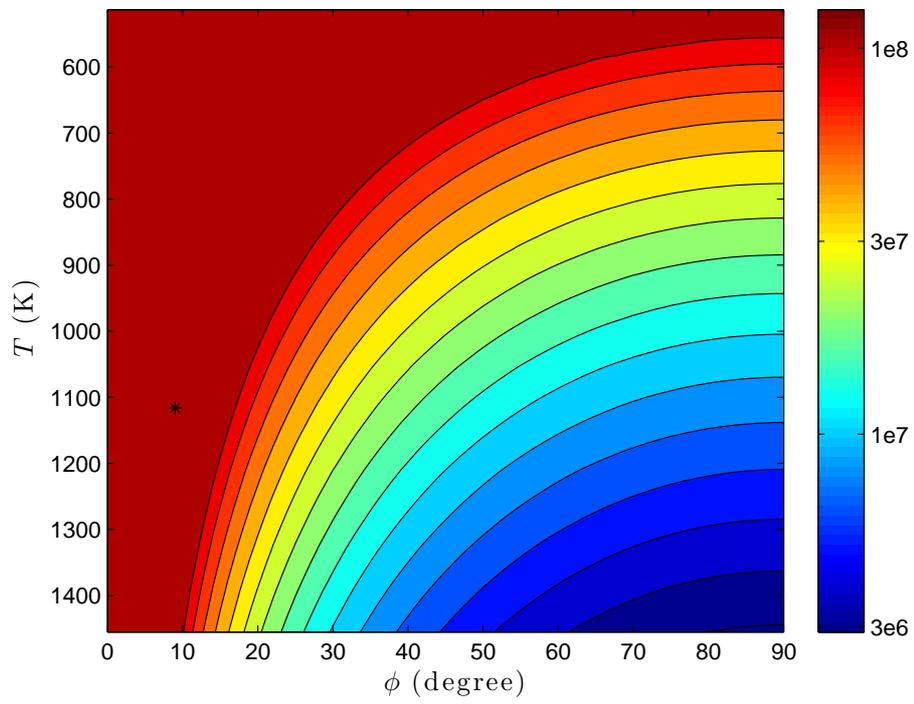}}
\caption{Profile of $K_{\rm eddy}$ (cm$^2$ s$^{-1}$) as a function of temperature $T$ (K) and latitude $\phi$ (degrees) for Jupiter. On the figure, we denote the location of CO quench level at 9$^{\circ}$ N where a measurement of its mixing ratio is available (B\'{e}zard et al. 2002). Our estimation gives $K_{\rm eddy} = (1.2\pm0.2)\times10^8$ cm$^2$ s$^{-1}$ at this location.} 
\label{fig: K_eddy_J}
\end{center}
\end{figure}

\begin{figure}
\begin{center}
\resizebox{\hsize}{!}{\includegraphics{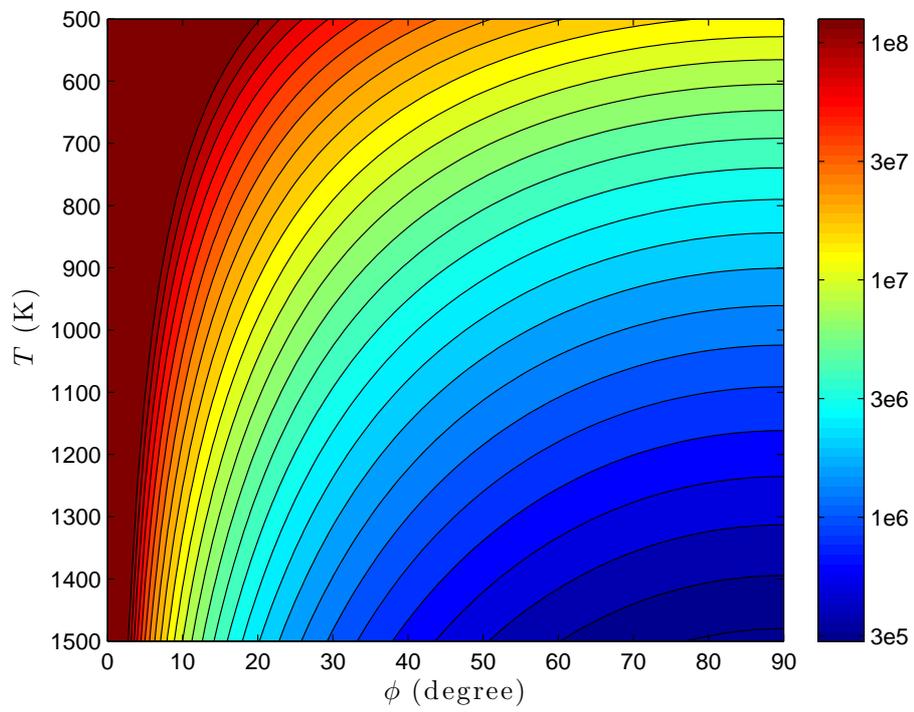}}
\caption{Profile of $K_{\rm eddy}$ (cm$^2$ s$^{-1}$) as a function of temperature $T$ (K) and latitude $\phi$ (degrees) for Saturn.}
\label{fig: K_eddy_S}
\end{center}
\end{figure}

Using equations (\ref{K_eddy_non}) and (\ref{K_eddy_rot}), we calculate  $K_{\rm eddy}$ as a function of temperature $T$ (radial direction) and latitude $\phi$ for both Jupiter and Saturn. The profile of $K_{\rm eddy}$ is shown in Fig \ref{fig: K_eddy_J} for Jupiter and Fig \ref{fig: K_eddy_S} for Saturn. A latitudinal dependence is clear, showing higher $K_{\rm eddy}$ near the equator where $\phi < \phi_t (T)$, and smaller $K_{\rm eddy}$ at higher latitudes where $\phi > \phi_t (T)$. The difference between equator and pole could be as large as one to two orders of magnitude. In Table 1, we compare the calculated $K_{\rm eddy}$ at a temperature level $T$ $\sim$ 1000 K (near the quench level) between Jupiter and Saturn. The values of $K_{\rm eddy}$ include uncertainties on the formulation itself and on the measured internal heat flux $F$. Near the equator, $K_{\rm eddy}$ is about $1\times10^8$ cm$^2$ s$^{-1}$ for both planets. For Jupiter, $K_{\rm eddy}$ decreases to about $1\times10^7$ cm$^2$ s$^{-1}$ at $\phi \sim 90^{\circ}$. For Saturn, $K_{\rm eddy}$ decreases to about $1\times10^6$ cm$^2$ s$^{-1}$ at $\phi \sim 90^{\circ}$.   

This latitudinal dependence of $K_{\rm eddy}$ is also shown in Flasar \& Gierasch (1977) and Fig 1 in Visscher et al. (2010). Both are based on the scalings given by equation (\ref{K_fg}) (Flasar \& Gierasch 1978). In section 2.2, we have shown that experimental results do not favor this scaling. In addition, previous studies are not able to determine the transition latitude $\phi_t$, and thus not able to calculate $K_{\rm eddy}$ for all latitudes. 

In summary, our new formulation of $K_{\rm eddy}$ is validated against experiments, thus providing a new perspective compared with previous theoretical investigations. Also, the pre-factors in the scalings are well determined by the experimental data, which enables us to constrain $K_{\rm eddy}$ much better than before.

\section{Deep Water Abundance Constrained by CO Thermochemistry Kinetics}
\label{CO_kinetics}

In this section, we update constraints on the deep water abundance of Jupiter and Saturn from CO using the newly constrained $K_{\rm eddy}$ in this paper. We used the rate limiting step proposed in Visscher \& Moses (2011), but we also considered a new CO kinetic model proposed in Venot et al. (2012), which has been applied to extrasolar planets' atmospheric chemistry (Venot et al. 2012) and Uranus atmospheric chemistry (Cavali\'{e} et al. 2014), but never before to Jupiter. We will show that these two kinetic models predict very different constraints on the deep water abundance. 

We make a few definitions regarding the abundance of species Z. The concentration of species Z is denoted as [Z] with a unit of molecules cm$^{-3}$. Mole fraction of Z is denoted as $X_{\rm Z}$ = [Z]/$n$, where $n$ is number density of the atmosphere (molecule cm$^{-3}$). Mixing ratio is denoted as $q_{\rm Z}$ = [Z]/[H$_2$]. The enrichment relative to solar is $E_{\rm Z}$ = $q_{\rm Z,planet}/q_{\rm Z,solar}$, where $q_{\rm Z,solar}$ is the mixing ratio of species Z in the Sun's atmosphere, taken from Asplund et al. (2009).    

\subsection{Constraints Using the Rate Limiting Step from Visscher $\&$ Moses (2011) }

We use a timescale approach, instead of solving the diffusion-kinetics equations explicitly. The timescale approach has been used extensively in previous studies to model the abundance of disequilibrium species, and it has been shown in Visscher et al. (2010) to be able to produce fairly accurate results. The error of the time-scale approach relative to the full diffusion-kinetics modeling in their particular example was $\sim$ 20$\%$. Since the relative error is acceptable in constraining Jupiter's deep water abundance, we choose to use the time-scale approach. Here is how we implement this approach. (1) We determine the chemical timescale $\tau_{\rm chem}$ along the adiabat; (2) we determine the mixing timescale $\tau_{\rm mix}$ along the adiabat using the newly constrained $K_{\rm eddy}$; (3) we equate $\tau_{\rm chem}$ and $\tau_{\rm mix}$ in order to find the quench level, and calculate the abundance of CO at the quench level. CO above quench level is vertically well mixed, so we can get the abundance of CO at a few bars as a function of $K_{\rm eddy}$ and the water abundance. Therefore, constraints can be put on the water abundance. Here we detail our implementation of this method to Jupiter's and Saturn's atmospheres using the rate limiting step from Visscher \& Moses (2011). 

\begin{figure}
\begin{center}
\resizebox{\hsize}{!}{\includegraphics{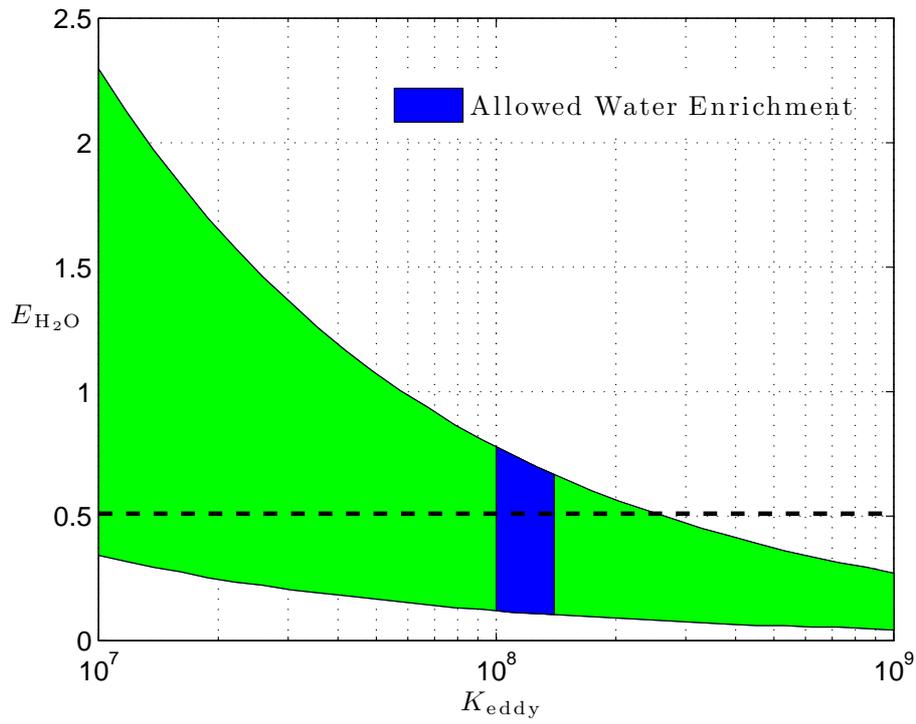}}
\caption{Chemical constraints on Jupiter's deep water abundance using rate limiting step from Visscher \& Moses (2011). The green area between two curves indicates the allowed $E_{\rm H_2O}$ considering a factor of five uncertainty in rate coefficient of rate limiting step (Jasper et al. 2007; Visscher \& Moses (2011)) and $X_{\rm CO} = (1.0\pm0.2)\times10^{-9}$ (B\'{e}zard et al. 2002). Using $K_{\rm eddy} = (1.2\pm0.2) \times 10^{8}$ cm$^2$ s$^{-1}$ constrained at the location of measurement, we find $E_{\rm H_2O} = 0.1 \sim 0.75$, corresponding to the blue area in the figure. Dashed line indicates $E_{\rm H_2O}$ measured by Galileo Entry Probe (Wong et al. 2004).}
\label{fig: CO_VM_J}
\end{center}
\end{figure}

\begin{figure}
\begin{center}
\resizebox{\hsize}{!}{\includegraphics{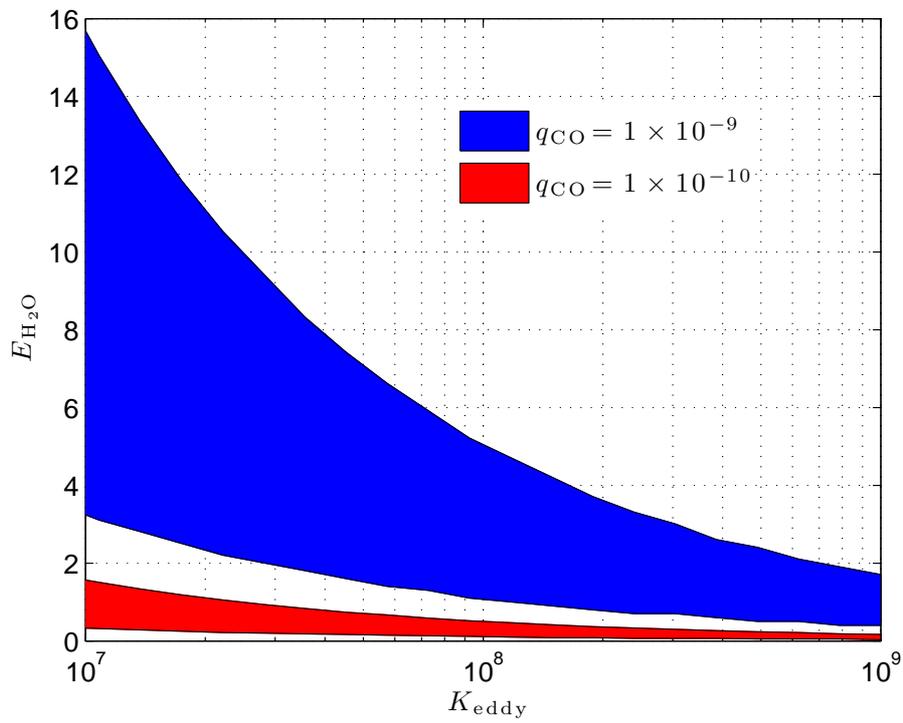}}
\caption{Chemical constraints on Saturn's deep water abundance using the the rate limiting step from Visscher \& Moses (2011). We consider two hypothetical tropospheric CO mixing ratio, namely, $q_{\rm CO}$ = $1\times10^{-9}$ and $q_{\rm CO}$ =$1\times10^{-10}$. The uncertainty indicated by shaded area is due to the factor of five uncertainty in the rate coefficient of the rate limiting step (Jasper et al. 2007; Visscher \& Moses 2011). We did not impose a constraint on $K_{\rm eddy}$ since it sensitively depends on the latitude where CO measurement is taken.}  
\label{fig: CO_VM_S}
\end{center}
\end{figure}

\begin{itemize}

\item{\it Chemical timescale $\bf{\tau_{\rm chem}}$}

We estimate $\tau_{\rm chem}$ using the rate limiting step proposed in Visscher \& Moses (2011): 
\begin{equation}
\label{CO_RLS}
\rm CH_3OH \, + \, M \, \leftrightarrow \, CH_3 \, + \,OH \, + \,M,
\end{equation}
where M represents any third body. 
The rate coefficients k$_{\rm \ref{CO_RLS}}$ for reaction (\ref{CO_RLS}) is calculated using the modified Tore parameters
in the Appendix of Jasper et al. (2007) (Typos in the appendix of that paper were corrected in author's website $\underline{http://www.sandia.gov/~ajasper/pub/}$). 

The chemical timescale $\tau_{\rm chem}$ can be expressed as
\begin{equation}\label{CO_timescale}
\tau_{\rm chem} (\textrm{CO}) \\=\\ \frac{X_{\rm CO}}{dX_{\rm CO}/dt}\ = \ \frac{X_{\rm CO}}{k_{\rm \ref{CO_RLS}}X_{\rm CH_3OH}n}. 
\end{equation}
In order to eliminate $X_{\rm CH_3OH}$ in equation ($\ref{CO_timescale}$), assuming an equilibrium state between CH$_3$OH and CO:
\begin{equation}\label{reaction:CH3OH}
\rm{CH_3OH} = \rm{CO}  +  \rm{2H_2},
\end{equation}
equation (\ref{CO_timescale}) can be rewritten as 
\begin{equation}
\tau_{\rm chem} (\textrm{CO})= \frac{1}{k_{\rm \ref{CO_RLS}}K_{\rm \ref{reaction:CH3OH},eq}X_{\rm H_2}^2 p^2 n},
\end{equation}
where the equilibrium constant K$_{\rm \ref{reaction:CH3OH},eq}$ is
\begin{equation}
K_{\rm \ref{reaction:CH3OH},eq} =
\textrm{exp} \left[-\frac{\Delta_f {G}_{\rm CH3OH} - \Delta_f {G}_{\rm CO}}{RT}\right].
\end{equation}
The Gibbs free energy of formation $\Delta_f {G}_{\rm CO}$ is from NIST-JANAF Thermochemical Table (Chase 1998), and $\Delta_f {G}_{\rm CH_3OH}$ is from CRC HandBook of Chemistry and Physics (Haynes et al. 2012).     

\item{\it Mixing timescale $\tau_{\rm mix}$}

The vertical mixing timescale can be expressed as
\begin{equation}
\tau_{\rm mix} = \frac{L_{\rm eff}^2}{K_{\rm eddy}},
\end{equation}
where $L_{\rm eff}$ is an effective length scale. It can be calculated following the recipe described in Smith (1998), and we find near the CO quench level, $L_{\rm eff}$ $\approx$ 0.12$H$ for Jupiter, and $L_{\rm eff}$ $\approx$ 0.14$H$ for Saturn. 

\item{\it $X_{\rm CO}$ at quench level}

We calculated $\tau_{\rm chem}$ and $\tau_{\rm mix}$  along Jupiter or Saturn's adiabat. The quench level was found by equating $\tau_{\rm chem}$ and $\tau_{\rm mix}$. The quench level depends on $K_{\rm eddy}$ and the reaction rate of the rate limiting step. For $K_{\rm eddy}$ = 1$\times$10$^8$ cm$^2$ s$^{-1}$, the quench temperatures for Jupiter and Saturn are about 1100 K and 1020 K, respectively. Once the quench level is determined, we can adjust the abundance of H$_2$O at the quench level to a value consistent with the observed CO abundance.

The equilibrium abundance of CO is governed by the net reaction (e.g. Fegley \& Prinn 1985, 1988; Lodders \& Fegley 2002; Visscher \& Fegley 2005)
\begin{equation}\label{CO equilibrium}
\rm{CO} + \rm{3H_2} = \rm{CH_4} + \rm{H_2O}.
\end{equation}
Using equilibrium constant K$_{\ref{CO equilibrium},\rm{eq}}$ of this reaction, the equilibrium abundance of CO can be expressed as 
\begin{equation}
X_{\rm CO} = \frac{K_{\ref{CO equilibrium},eq}X_{\rm CH_4}X_{\rm H2O}}{X_{\rm H_2}^3 p^2},
\end{equation}
where $p$ is the atmospheric pressure in the unit of bars. On Jupiter, the mole fractions of H$_2$ and CH$_4$ were measured by Galileo Probe Mass Spectrometer (GPMS), with a value of $X_{\rm H_2}$ = 0.864 $\pm$ 0.003 (Niemann et al. 1998), and $X_{\rm CH_4}$ = (2.05 $\pm$ 0.32) $\times$ 10$^{-3}$ (Wong et al. 2004). For Saturn, CH$_4$ is measured by Cassini CIRS with a mixing ratio $q_{\rm CH_4}$ = (5.3$\pm$0.2)$\times$10$^{-3}$ (Flasar et al. 2005; Fletcher et al. 2009a).
     
The equilibrium constant K$_{\ref{CO equilibrium},\rm{eq}}$ can be calculated as  
\begin{equation}
K_{\rm \ref{CO equilibrium},eq}= \textrm{exp} \left [-\frac{\Delta_f  {G}_{\rm CO} - \Delta_f {G}_{\rm
CH_4} - \Delta_f {G}_{\rm H_2O}}{ RT } \right ],
\end{equation}
where $R$ = 8.314 J mol$^{-1}$ K$^{-1}$ is the universal gas constant, and $\Delta_f {G}_{\rm CO}$, $\Delta_f {G}_{\rm
CH_4}$, $\Delta_f {G}_{\rm H_2O}$ are Gibbs free energy of formation for CO, $\rm CH_4$, and $\rm H_2O$, respectively. We took these values from NIST-JANAF Thermochemical Tables (Chase 1998). 

\item{\it Constraints on $X_{\rm H_2O}$ for Jupiter and Saturn}

Now that we can calculate $X_{\rm CO}$ as a function of $K_{\rm eddy}$ and $X_{\rm H_2O}$ using the time-scale approach, given the observed CO abundance in the troposphere, we can place constraints on $X_{\rm H_2O}$.

On Jupiter, the tropospheric CO was measured by B\'{e}zard et al. (2002) near one of the hot spots at 9$^{\circ}$ N. $X_{\rm CO}$ was estimated to be (1.0 $\pm$ 0.2)$\times$10$^{-9}$ at the 6 bar level. In Figure \ref{fig: CO_VM_J}, we plot the allowed $E_{\rm H_2O}$ (enrichment relative to solar $q_{\rm O}$ = 9.8$\times$10$^{-4}$ from Asplund et al. 2009) as a function of $K_{\rm eddy}$. We consider a factor of five uncertainty in the rate limiting step. Near 9$^{\circ}$ N, $K_{\rm eddy} =(1.2\pm0.2)\times10^8$ cm$^2$  s$^{-1}$ is found according to Table 1. Applying this constraint on $K_{\rm eddy}$, we find $E_{\rm H_2O}$ = 0.1 $\sim$ 0.75. This constraint is consistent with the Galileo measurement of $E_{\rm H_2O}$ = 0.50 $\pm$ 0.16 (Wong et al. 2004). 

On Saturn, CO is observed at a mole fraction of (1.5$\pm$0.8)$\times$10$^{-9}$ (Noll et al. 1986; Noll \& Larson 1991). However, the fraction coming from an internal source (vertical mixing) is still unknown. Cavali\'{e} et al. (2009) put an upper limit on the amount of CO from an internal source with $q_{\rm CO}< 1\times10^{-9}$, however, no lower limit is obtained yet. Here we keep tropospheric CO abundance as a free parameter and explore two cases, namely, $q_{\rm CO}$ = 1.0$\times$10$^{-9}$ and $q_{\rm CO}$ = 1.0$\times$10$^{-10}$. The constraints on the deep water abundance are shown in Fig \ref{fig: CO_VM_S}. Once the tropospheric CO is measured in the future, we can refer to Fig \ref{fig: K_eddy_S} or Table \ref{tab: Comparison} to find out the corresponding $K_{\rm eddy}$ at the observation location, then refer to Fig. 6 to find out the constraints on Saturn's $X_{\rm H_2O}$. 
\end{itemize}

\subsection{Constraints Using Kinetic Model from Venot et al. (2012)}

Venot et al. (2012) proposed a carbon-nitrogen kinetic model and applied it to study hot Jupiter atmospheres. The kinetic model was originally developed for modeling combustion process in car engines and has been validated at a range of temperatures from 300 K to 2500 K, and pressure from 0.01 bar to some hundred bars (Venot et al. 2012). Considering the relevant range of temperature and pressure, this kinetic model is appropriate to study Jupiter and Saturn's disequilibrium chemistry as well. It would be useful to compare the implied $X_{\rm H_2O}$ from Visscher \& Moses (2011)'s kinetic model (\textit{VM model}) and Venot et al. (2012)'s kinetic model (\textit{Venot model}) . To our knowledge, this comparison has never been done for Jupiter and Saturn.

Since no rate limiting step has been identified from \textit{Venot model}, the chemical timescale cannot be explicitly calculated. Therefore, we developed a full diffusion-kinetic model, incorporating the whole kinetic network from Venot et al. (2012). Our model is similar to the model developed in Venot et al. (2012), but with application to Jupiter's and Saturn's atmospheres. For each species $i$, we solve the diffusion-kinetic equation 
\begin{equation}
\frac{\partial{n_i}}{\partial{t}} + \frac{\partial{\phi_i}}{\partial{z}} = S_i,
\end{equation}  
where $n_i$ is the number density of species $i$, $S_i$ is the net production(loss) rate, and $\phi_i$ is the vertical flux given by 
\begin{equation}
\phi_i = n_i K_{\rm eddy}\frac{1}{y_i} \frac{\partial{y_i}}{\partial{z}},
\end{equation}
where $y_i$ is the mass fraction of species i. The chemical source term $S_i$ is calculated by \textit{cantera}, a software toolkit for solving problems involving chemical kinetics (Goodwin  et al. 2014). We take the equilibrium state as the initial condition of our model, and let the system evolve into a steady state. The steady state $X_{\rm CO}$ can be extracted from our model, for given inputs of $K_{\rm eddy}$ and $E_{\rm H2O}$.  

We check the consistency between our model and the diffusion-kinetic model used in Mousis et al. (2014), the latter of which is adapted from the model used in Venot et al. (2012). Assuming O/H is 21 times solar, C/H is 9 times solar, and $K_{\rm eddy}$ = 10$^9$ cm$^2$ s$^{-1}$, the model in Mousis et al. (2014) derived a CO mole fraction of 1$\times$10$^{-9}$ for Saturn. Using the same input, our model derived a CO mole fraction of 1.1$\times$10$^{-9}$. Therefore, the two models are producing similar results, with the remaining differences probably coming from the uncertainties on the adiabat. 

Although the diffusion-kinetic model is robust in calculating $X_{\rm CO}$ for given $K_{\rm eddy}$ and $E_{\rm H_2O}$, it is very time-consuming to explore the parameter space of ($K_{\rm eddy}$, $E_{\rm H_2O}$, $X_{\rm CO}$) using this method. Therefore, we developed an approximation method based on the timescale approach, which is easy and quick to implement, and accurate enough compared with the full diffusion-kinetic modeling method. 
The timescale approach requires the estimation of the chemistry timescale $\tau_{\rm chem}$. We assume there exists a rate limiting step $\rm{R1} + \rm{R2} \rightarrow \rm{P1}+\rm{P2}$, then 
\begin{equation}
\tau_{\rm chem}  = \frac{[\rm{CO}]}{d[\rm{CO}]/dt} = \frac{[\rm{CO}]}{k_{rls}[\rm{R1}] [\rm{R2}]}, 
\end{equation}
where $k_{rls}$ is the rate coefficient of the rate limiting step. Note that $k_{rls}$ is generally proportional to $e^{-E/T}$, where $E$ is the activation energy. [R1] and [R2] can be related to $[\rm{CO}]$ and $[\rm{H_2}]$ via the equilibrium constants and some powers of pressure. The equilibrium constants are proportional to $e^{\Delta G/T}$, where $\Delta G$ is the change of Gibbs free energy of formation in the reaction. Therefore, it is reasonable to assume $\tau_{\rm chem} = Ce^{A/T}p^{\alpha}$, where $A$, $C$ and $\alpha$ are constant coefficients that need to be fitted. 

We fit the coefficients based on the numerical results from our diffusion-kinetic model. Note that for each diffusion-kinetic simulation, we have numerical values for ($K_{\rm eddy}$, $E_{\rm H_2O}$, $X_{\rm CO}$), where $K_{\rm eddy}$ and $E_{\rm H_2O}$ are input parameters, and $X_{\rm CO}$ is the simulation result. We use the following procedure to fit the coefficients $A$, $C$ and $\alpha$. (1) With $X_{\rm CO}$ and $E_{\rm H_2O}$, we used equilibrium chemistry of CO to find out the quench temperature $T_{q}$ and quench pressure $p_{q}$. (2) With $K_{\rm eddy}$, we calculated the mixing timescale at the quench level. The chemical timescale at quench level $\tau_{\rm chem,q}$ is equal to the mixing timescale. (3) We therefore have one set of numerical values for ($T_q$, $p_q$, $\tau_{\rm chem,q}$) from each simulation. We run three simulations for Jupiter and three simulations for Saturn with different $K_{\rm eddy}$ and $E_{\rm H_2O}$, and use the six sets of data to fit the coefficients A, C and $\alpha$. 
The derived fitting formula for $\tau_{\rm chem}$ is: 
\begin{equation}\label{tau_chem_venot}
\tau_{\rm chem} \approx (5\times10^{-6} e^{2.8\times10^4 /T})p^{-1.29} \ \textrm{s}.
\end{equation}    
With the chemical timescale $\tau_{\rm chem}$ available, we followed the procedures described in Section 3.1 to implement the timescale approach. Note that the effective length scale $L_{\rm eff}$ is calculated following the recipe described in Smith (1998), which has dependence on $\tau_{\rm chem}$. For the \textit{Venot model}, we find $L_{\rm eff}\approx 0.10H$ for Jupiter, and $L_{\rm eff} \approx 0.12H$ for Saturn. We compared $X_{\rm CO}$ calculated by the timescale approach with $X_{\rm CO}$ calculated by the full-diffusion kinetic modeling in Table 2. For different combinations of $K_{\rm eddy}$ and $E_{\rm H_2O}$, we find the difference on $X_{\rm CO}$ is within 10$\%$ between the timescale approach and full-diffusion kinetic modeling. This validated the timescale approach and the $\tau_{\rm chem}$ we estimated in equation (\ref{tau_chem_venot}). 

\begin{figure}
\begin{center}
\resizebox{\hsize}{!}{\includegraphics{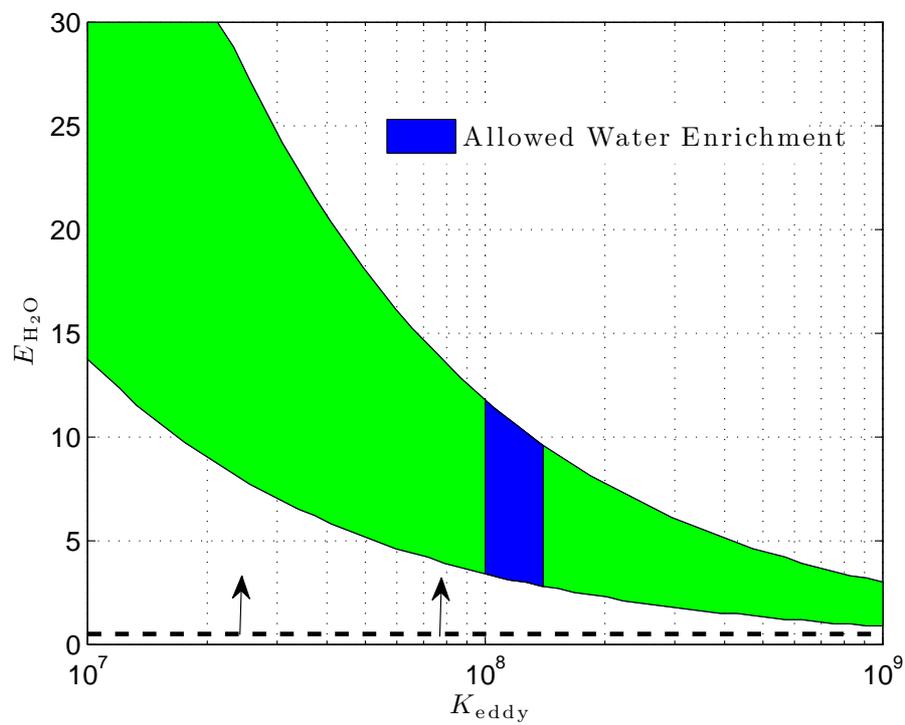}}
\caption{Chemical constraints on Jupiter's deep water abundance using the CO kinetic model from Venot et al. (2012). The green area indicates allowed $E_{\rm H_2O}$ considering a factor of two uncertainty in the rate coefficient and $X_{\rm CO} = (1.0\pm0.2)\times10^{-9}$ (B\'{e}zard et al. 2002). Using $K_{\rm eddy} = (1.2\pm0.2)\times10^8$ near 9$^{\circ}$ N, we find $E_{\rm H_2O} = 3 \sim 11$, corresponding to the blue area in the figure.}
\label{fig: CO_Venot_J}
\end{center}
\end{figure}

\begin{figure}
\begin{center}
\resizebox{\hsize}{!}{\includegraphics{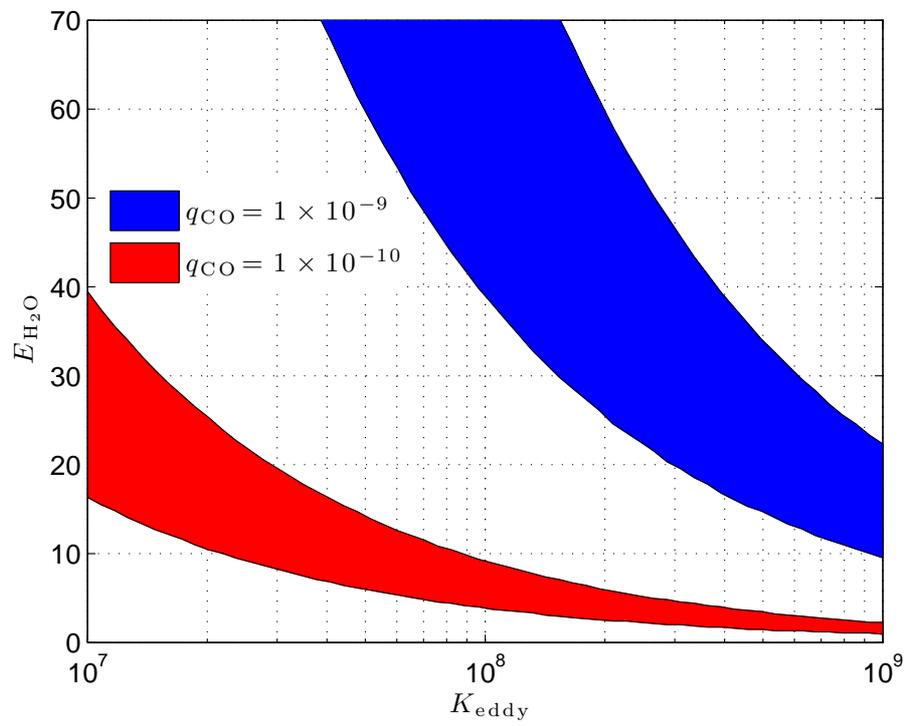}}
\caption{Chemical constraints on Saturn's deep water abundance using CO kinetic model from Venot et al. (2012). We consider two hypothetical tropospheric CO mixing ratio, namely, $q_{\rm CO}$ = $1\times10^{-9}$ and $q_{\rm CO}$ =$1\times10^{-10}$. The uncertainty indicated by shaded area is due to the uncertainty of the rate coefficients. We did not impose a constraint on $K_{\rm eddy}$ since it sensitively depends on the latitude where CO measurement is taken.} 
\label{fig: CO_Venot_S}
\end{center}
\end{figure}

Using the timescale approach, we derived the constraint on Jupiter's $E_{\rm H_2O}$. The constraint is shown in Fig \ref{fig: CO_Venot_J} where find $E_{\rm H_2O} = 3\sim11$. As a comparison, $E_{\rm H_2O} = 0.1 \sim 0.75$ using the \textit{VM model}. 
In Fig. \ref{fig: CO_Venot_S}, we plot the constraint on $E_{\rm H_2O}$ for Saturn using the \textit{Venot model}. We did not constrain $K_{\rm eddy}$ here since it sensitively depends on latitude. In the future, once the abundance of tropospheric CO is measured, we can refer to Fig \ref{fig: K_eddy_S} or Table \ref{tab: Comparison} to get $K_{\rm eddy}$, then refer to Fig. \ref{fig: CO_Venot_S} to find the constraint on $E_{\rm H_2O}$. 

\section{Discussion}
\label{dis}

In this paper, we revisited the constraints on Jupiter's deep water abundance by disequilibrium species CO. We proposed a new formulation of eddy diffusion coefficient, based on laboratory studies of turbulent rotating convection. With newer eddy diffusion coefficient, we updated the constraints on Jupiter's deep water abundance. Using the rate limiting step from Visscher \& Moses (2011), we find $E_{\rm H_2O} = 0.1 \sim 0.75$. We also consider another chemical model from Venot et al. (2012), and the constraints on deep water abundance are $E_{\rm H_2O} = 3 \sim 11$. We do not consider the possibility of a strong compositional stratification of either Jupiter or Saturn in which the heavy element abundance increases toward the center of the body (Leconte \& Chabrier 2012). It is possible that the gravitational field measurements to be made by Juno at Jupiter and Cassini at Saturn will provide constraints on the degree of such differentiation, but the problem of deriving oxygen abundance from such differentiation may be difficult to resolve.

The distinct ranges of {\it E}$_{\rm H_2O}$ found in Jupiter with the two kinetic models require very different formation conditions. The range of {\it E}$_{\rm H_2O}$ ($\sim$0.1--0.75) found with the kinetic model of Visscher \& Moses (2011) necessities the formation of Jupiter in a region of the protosolar nebula that is strongly depleted in oxygen. A moderate H$_2$O enrichment by a factor of $\sim$2 in Jupiter already corresponds to a substantial \textit{depletion} of the oxygen abundance by a factor of $\sim$2 in its feeding zone (Mousis et al. 2012). This implies that the oxygen abundance in Jupiter's feeding zone should be depleted by factors of $\sim$5--40 times, compared to the protosolar abundance, for values of {\it E}$_{\rm H_2O}$ found in the $\sim$0.1--0.75 range in the envelope. Such a high oxygen depletion might be explained if Jupiter formed at a slightly lower heliocentric distance than the iceline in the protosolar nebula. At this location, the diffusive redistribution and condensation of water vapor induces two effects: it increases the density of ice at the position of the iceline but it also drops the water vapor abundance at distances slightly closer to the Sun (Stevenson \& Lunine 1988; Ali-Dib et al. 2014). However, at this location, because the disk's temperature is higher than the water condensation temperature, it becomes difficult to accrete efficiently icy planetesimals in Jupiter's envelope in order to explain the giant planet's overall elevated metallicity. In this context, a possible explanation of the observed enrichments in Jupiter could lie in its late formation in the protosolar nebula. In this case, the photoevaporation of the disk and the delivery of condensible species in vapor forms from its outer regions may lead to a progressive homogeneous enrichment of the disk in heavy elements (Guillot \& Hueso 2006). Jupiter's metallicity would be then representative of the heavy element enrichment acquired by the disk's gas from which it accreted.

On the other hand, the range of {\it E}$_{\rm H_2O}$ ($\sim$3--11) found with the kinetic model of Venot et al. (2012) corresponds to cases of Jupiter formation in environments where the O abundance varies from moderately depleted to slightly enriched compared to the protosolar value. The value of {\it E}$_{\rm H_2O}$ $\sim$7 is predicted in Jupiter when it accreted planetesimals formed from a gas phase of protosolar composition (Mousis et al. 2012). In this case, Jupiter's building blocks were agglomerated from a mixture of clathrates and pure ices condensed down to $\sim$22 K in the protosolar nebula. Any value of {\it E}$_{\rm H_2O}$ lower than $\sim$7 requires the formation of Jupiter at a slightly lower heliocentric distance than the ice line in the protosolar nebula. In contrast, values of {\it E}$_{\rm H_2O}$ higher than $\sim$7 correspond to an increase of the water abundance in the giant planet's feeding zone, thus easing the trapping of volatiles in the form of clathrates at higher disk's temperature (in the $\sim$50--80 K temperature range; Mousis et al. 2009, 2012). In both cases, the volatiles responsible for the enrichments measured at Jupiter were supplied either via the partial erosion of its core or via accretion of planetesimals dragged from the nebula during the hydrodynamical collapse of the envelope.

In the case of Saturn, the range of {\it E}$_{\rm H_2O}$ remains still loosely constrained because no inner limit has been found for the internal source of CO. For the moment, depending on the value of {\it E}$_{\rm H_2O}$, the range of conclusions made for Jupiter applies to Saturn as well. When an inner limit for the internal source of CO is set in the future, it will be possible to derive more specific conclusions. 

\begin{figure}
\begin{center}
\resizebox{\hsize}{!}{\includegraphics{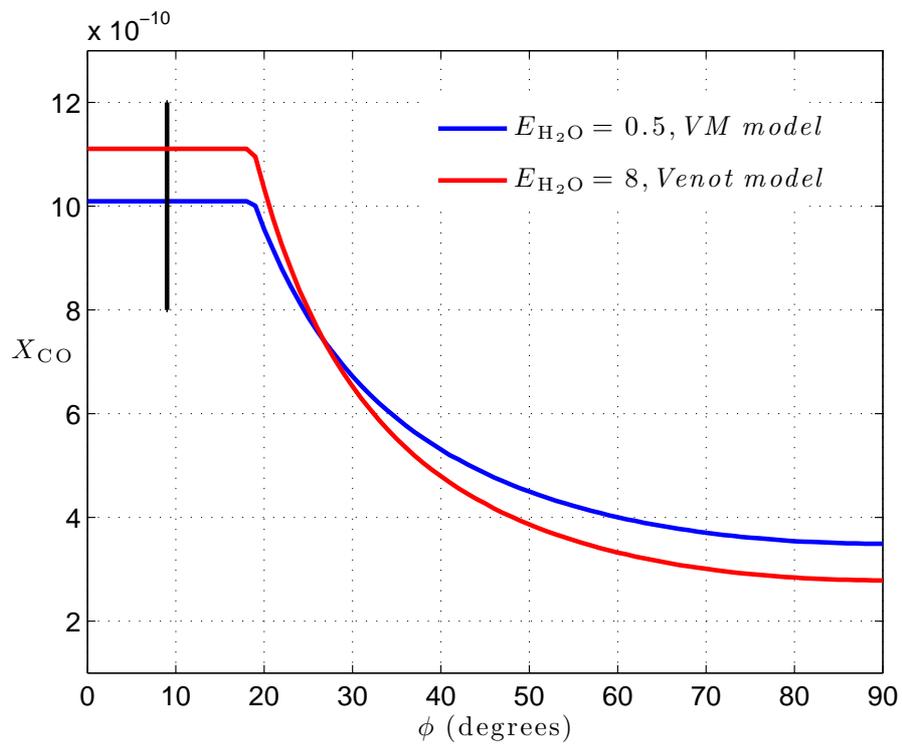}}
\caption{Prediction of tropospheric $X_{\rm CO}$ as a function of latitude for a given $E_{\rm H_2O}$. We consider two CO kinetic models, namely, the \textit{VM model} (Visscher \& Moses 2011) and the \textit{Venot model} (Venot et al. 2012). The error bar indicates the measurement of $X_{\rm CO}$ at 9 degrees by B\'{e}zard et al. (2002)}
\label{fig: CO_horizontal}
\end{center}
\end{figure}

\begin{figure}
\begin{center}
\resizebox{\hsize}{!}{\includegraphics{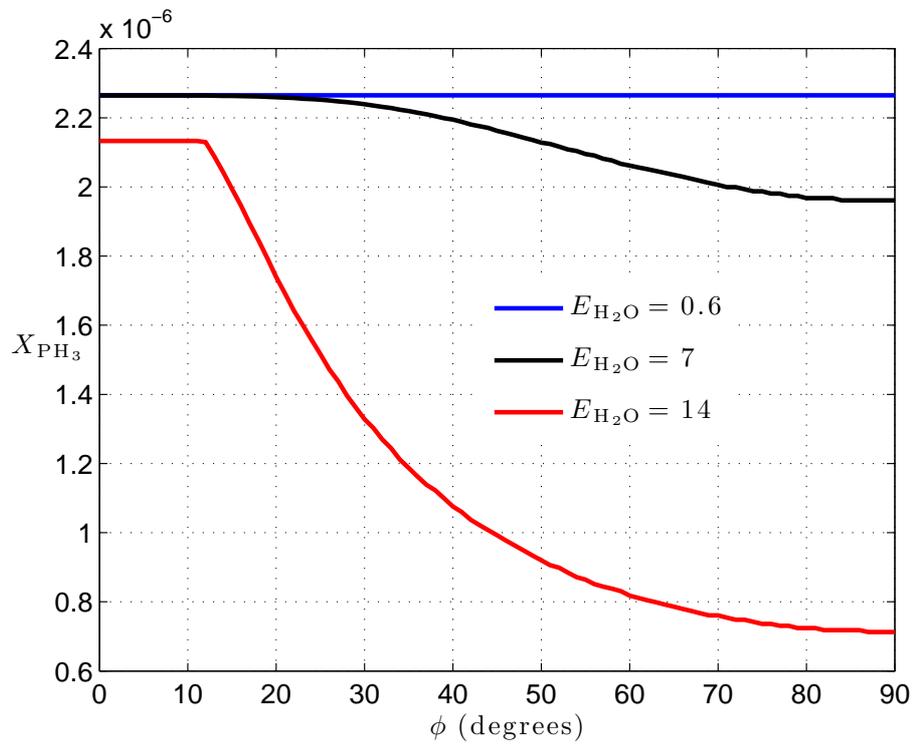}}
\caption{Prediction of $X_{\rm PH_3}$ as a function of latitudes for given values of $E_{\rm H_2O}$. The bulk phosphorus abundance is taken as $X_{\rm P}$ = 
2.3$\times$10$^{-6}$ (Mousis et al. 2012). The rate limiting step for PH$_3$ - P$_4$O$_6$ conversion is taken as equation (\ref{PH3_RLS}) (Visscher \& Fegley 2005).}
\label{fig: PH3_horizontal}
\end{center}
\end{figure}

Our model is still subject to improvements in the following aspects:  
\begin{itemize}

\item{\it CO kinetic models}

With a better assessment of the kinetics of chemical reactions, our model should allow derivation of a much narrower range of deep water abundances in Jupiter, and subsequently provide more robust constraints on their formation conditions. Currently, the two kinetic models, \textit{VM model} and \textit{Venot model}, place very different constraints on deep water abundance. The \textit{VM model} is derived from previous Jupiter and Saturn models (e.g., Gladstone et al. 1996; Moses et al. 1995a,b, 2000a,b), with extensive updates on high temperature kinetics from combustion chemistry studies (Visscher et al. 2010). As a comparison, the \textit{Venot model} is based on a C0-C2 reaction base originally developed for industrial applications. A mechanism for nitrogen is coupled to the C0-C2 reaction base to model C/N/O/H chemistry. According to Venot et al. (2012), their model has been validated by various experiments over a large range of pressure and temperature (e.g., Battin-Leclerc et al. 2006; Bounaceur et al. 2007; Anderlohr et al. 2010; Bounaceur et al. 2010; Wang et al. 2010), while the \textit{VM model} has not been validated against experiments. From this aspect, \textit{Venot model} is more plausible than the \textit{VM model}. 

\item{\it Potential Tests on $K_{\rm eddy}$ Formulation Using JIRAM on Juno}

Although our new formulation of $K_{\rm eddy}$ has been validated against laboratory experiments, there are still no observational constraints on Jupiter's or Saturn's $K_{\rm eddy}$. 

The microwave radiometer onboard Juno spacecraft should be able to measure the deep water abundance in Jupiter (Janssen et al. 2005). If performed, this measurement should be able to place constraints on $K_{\rm eddy}$, as is shown in Fig. \ref{fig: CO_VM_J} and Fig. \ref{fig: CO_Venot_J}. 

We have predicted the dependence of $K_{\rm eddy}$ on latitude, thus the concentration of disequilibrium species should have a latitudinal variation. Measurement of disequilibrium species at different latitudes will be able to provide a test of the latitudinal dependence. In Fig. \ref{fig: CO_horizontal}, we plot the calculated $X_{\rm CO}$ in the troposphere as a function of latitudes. $X_{\rm CO}$ near the equator is about three times $X_{\rm CO}$ near the pole. Currently the only measurement of tropospheric CO is at 9$^{\circ}$ N (B\'{e}zard et al. 2002). Measurement of CO at higher latitudes with less than 20$\%$ uncertainty would be able to distinguish the latitudinal variation, and thus test our formulation of $K_{\rm eddy}$.  

Another disequilibrium species that has been detected in Jupiter's atmosphere is PH$_3$. Tropospheric abundance of PH$_3$ was measured by Cassini CIRS. The average value of $X_{\rm PH_3}$ below 1 bar is estimated to be $(1.9\pm0.1) \times 10^{-6}$ (Irwin et al. 2004; Fletcher et al. 2009b), which is order of magnitude larger than its equilibrium abundance, indicating its state of disequilibrium. The net reaction for $\rm {PH_3}$ destruction is (Prinn et al. 1984; Fegley \& Prinn 1985; Visscher \& Fegley 2005)
\begin{equation}\label{PH3 equilibrium}
4\rm{PH_3} + 6\rm{H_2O} = \rm{P_4O_6} + 12\rm{H_2}.
\end{equation}  
Following Visscher \& Fegley (2005), we use the reaction 
\begin{equation}\label{PH3_RLS}
\rm{PH_3} + \rm{OH} \rightarrow \rm{H_2POH} + \rm{H}
\end{equation} 
as the rate limiting step. The bulk phosphorus abundance is unknown. We take $X_{P} = 2.3\times10^{-6}$, which is consistent with the calculations by Mousis et al. (2012). In Fig. \ref{fig: PH3_horizontal}, we plot the calculated $X_{\rm PH_3}$ as a function of latitude. The profile sensitively depends on the deep water abundance. For low water abundance (e.g. $E_{\rm H_2O}$ = 0.6), $\rm PH_3$ is the dominant species regardless of $K_{\rm eddy}$. However, at large water abundance (e.g. $E_{\rm H_2O}$ = 14), the dependence on $K_{\rm eddy}$ is very sensitive. 

The spectrometer JIRAM on board Juno will be able to retrieve the abundance of CO and PH$_3$ at 3$\sim$10 bars. The relative error for CO should be about 60$\%$, and for PH$_3$ it should be about 30$\%$ (Grassi et al. 2010). From Fig. \ref{fig: CO_horizontal}, the latitudinal variation of CO is at the limit of instrument (60$\%$ relative error). From Fig. \ref{fig: PH3_horizontal}, if Jupiter has a bulk water abundance of $E_{\rm H_2O} \gtrsim 7$, then JIRAM should be able to see the latitudinal variation, otherwise, the latitudinal variation is too small to be resolved by JIRAM. 
 
\item{\it Effects of Horizontal Mixing}

Although we predicted a latitudinal variation of CO and PH$_3$, we ignored the effect of possible horizontal mixing that tends to homogenize latitudinal gradients. If we assume horizontal mixing is driven by eddy diffusion, and assume horizontal $K_{\rm eddy}$ is of similar order to vertical $K_{\rm eddy}$, then the horizontal mixing timescale $\tau_{\rm mix,h} \sim R_J^2/K_{\rm eddy} \sim 10^{11}$ s, which is much larger than the vertical mixing timescale $\tau_{\rm mix} \sim 10^6$ s. Therefore, horizontal eddy diffusion is not able to effectively homogenize disequilibrium species in the troposphere. However, if significant horizontal circulation across latitudes exists, the horizontal mixing could be enhanced. Consider the meridional velocity $v$, and meridional scale $R_J$, then the horizontal mixing timescale $\tau_{\rm mix,h} \sim R_J/v$. For $v$ = 10 m s$^{-1}$, $\tau_{\rm mix,h} \sim 7\times10^6$ s. Compared with vertical mixing timescale $\tau_{\rm mix} \sim 10^6$ s, horizontal mixing could smooth out the latitudinal gradient to some degrees. The determination of horizontal profile of CO or PH$_3$ under horizontal mixing would require the knowledge of tropospheric circulation of Jupiter, which is still unknown.  

Although we do not know exactly the extent of horizontal mixing, we can explore its effect on the constraints of $E_{\rm H_2O}$ by considering two extreme conditions, namely, no mixing and 100$\%$ mixing. For no mixing, the results are $E_{\rm H_2O} = 0.1 \sim 0.75$ using the \textit{VM model} and $E_{\rm H_2O} = 3 \sim 11$ using the \textit{Venot model}. For 100$\%$ mixing, we assume CO is well mixed across latitudes. According to our calculations, the results are $E_{\rm H_2O} = 0.2 \sim 1.0$ using the \textit{VM model} and $E_{\rm H_2O} = 4.7 \sim 16.3$ using the \textit{Venot model}. Therefore, horizontal mixing does not significantly affect our constraints on Jupiter's deep water abundance. 

\end{itemize}

\section{Conclusion}
In this paper, we improved the thermochemical constraints on Jupiter's deep water abundance in two aspects. First, we developed a new formulation for eddy diffusion coefficient based on experiments dedicated to turbulent rotating convection. Application of the new formula to Jupiter and Saturn reveals a smooth transition from slow rotation regime (near the equator) to rapid rotation regime (near the pole), and a strong latitudinal dependence. We estimate an uncertainty for our newly-derived coefficient of less than 25$\%$, which is much better than the one order of magnitude used in the literature. Secondly, we considered two updated chemical-kinetic models and derived the constraints on Jupiter's deep water abundance. Using the rate limiting step proposed by Visscher $\&$ Moses (2011), we find the enrichment of water (relative to solar) for Jupiter is $E_{\rm H_2O} = 0.1\sim0.75$, while using the chemical-kinetic model proposed by Venot et al. (2012), we find $E_{\rm H_2O} = 3\sim11$.  With a better assessment of chemical kinetics, our model should allow deriving a much narrower range of deep water abundance in Jupiter. The constraint on Saturn's deep water abundance is still loose due to the lack of measurements of tropospheric CO abundance.   
    
\section*{Acknowledgments}
We thank Dr. Thibault Cavali\'{e} and an anonymous referee for their useful comments that helped us improve our manuscript.   
We thank Olivier Desjardins for helpful discussions. J. I. L. is grateful for support from the Juno Project during the conduct of this work. O. M. acknowledges support from CNES. Part of this work has been carried out thanks to the support of the A*MIDEX project (n\textsuperscript{o} ANR-11-IDEX-0001-02) funded by the ``Investissements d'Avenir'' French Government program, managed by the French National Research Agency (ANR).





\begin{table}[h]
\centering \caption{Comparison between timescale approach and full diffusion-kinetic modeling using the \textit{Venot model}.}
\begin{center}
\begin{tabular}{lcc}
\hline \hline
($K_{\rm eddy}$, $E_{\rm H_2O}$)$^a$ & $X_{\rm CO}$ from approximation method & $X_{\rm CO}$ from full diffusion-kinetic modeling \\
\hline
Jupiter & & \\
($1\times10^7$, 20) &   $7.6\times10^{-10}$   &  $7.5\times10^{-10}$    \\
($1\times10^8$, 10)  &  $1.5\times10^{-9}$ &  $1.5\times10^{-9}$    \\
($1\times10^9$, 1.0)  &  $6.0\times10^{-10}$ &  $5.7\times10^{-10}$    \\
\hline
Saturn & & \\
($1\times10^7$, 20) &   $7\times10^{-11}$   &  $6.8\times10^{-11}$    \\
($1\times10^8$, 10)  &  $1.5\times10^{-10}$ &  $1.6\times10^{-10}$    \\
($1\times10^9$, 1.0)  &  $6.0\times10^{-11}$ &  $6.5\times10^{-11}$    \\
\multicolumn{3}{l}{$^a$\footnotesize{The unit of $K_{\rm eddy}$ is cm$^2$ s$^{-1}$}}
\end{tabular}
\end{center}
\label{tab: Comparison_H2O}
\end{table}

\clearpage

\clearpage

\clearpage

\clearpage

\clearpage

\clearpage

\clearpage

\clearpage

\clearpage

\clearpage

\end{document}